\documentclass[sigplan,10pt]{acmart}
\renewcommand\footnotetextcopyrightpermission[1]{}
\usepackage{amsmath}
\usepackage{booktabs}
\usepackage{multirow}
\usepackage{algorithm}
\usepackage{algpseudocode}
\usepackage{subcaption}
\usepackage{xspace}
\usepackage{url}

\begin{document}
\pagestyle{plain}

\title{LifeTrain: Training-State Lifecycle Scheduling for Large Language Model Training on Bandwidth-Constrained Heterogeneous Supercomputers}

\author{Yao Lu}
\affiliation{%
\institution{Sino-German Joint Software Institute, Beihang University}
\city{Beijing}
\country{China}
}
\email{[luyuan@buaa.edu.cn](mailto:luyuan@buaa.edu.cn)}

\author{Shiqing Ma}
\affiliation{%
\institution{Sino-German Joint Software Institute, Beihang University}
\city{Beijing}
\country{China}
}

\author{Zhongzhi Luan}
\affiliation{%
\institution{Sino-German Joint Software Institute, Beihang University}
\city{Beijing}
\country{China}
}
\email{[luan.zhongzhi@buaa.edu.cn](mailto:luan.zhongzhi@buaa.edu.cn)}

\author{Gen Li}
\affiliation{%
\institution{Sino-German Joint Software Institute, Beihang University}
\city{Beijing}
\country{China}
}

\author{Jiaxing Qi}
\affiliation{%
\institution{Sino-German Joint Software Institute, Beihang University}
\city{Beijing}
\country{China}
}

\author{Bin Han}
\affiliation{%
\institution{Sino-German Joint Software Institute, Beihang University}
\city{Beijing}
\country{China}
}

\author{Hailong Yang}
\affiliation{%
\institution{Sino-German Joint Software Institute, Beihang University}
\city{Beijing}
\country{China}
}

\author{Depei Qian}
\affiliation{%
\institution{Sino-German Joint Software Institute, Beihang University}
\city{Beijing}
\country{China}
}

\begin{abstract}
Production heterogeneous supercomputing platforms are increasingly used to host large language model (LLM) training workloads. However, existing GPU-oriented training runtimes typically rely on high-bandwidth device memory, fast interconnects, and mature collective communication libraries, making them difficult to directly adapt to MT-3000, a platform with an explicit memory hierarchy, limited usable DDR capacity, and constrained inter-cluster communication. This paper presents RATrain, a resource-aware training runtime for dense LLMs on bandwidth-constrained heterogeneous supercomputing platforms. RATrain formulates standard non-interleaved 1F1B training as a training-state lifecycle scheduling problem, and schedules gradient synchronization, parameter update, parameter-view prefetching, and activation recovery at layer-level and stage-local granularity. RATrain further combines an MT-3000-aware execution backend for efficient and predictable FP16 GEMM, Attention Backward, and explicit data movement with a resource-aware planner that selects feasible training configurations under the 20GB usable-DDR constraint per compute cluster. We implement RATrain on a real MT-3000 platform and evaluate it using LLaMA-2-7B, Baichuan2-13B, Qwen2.5-32B, and LLaMA-2-70B configurations. Results show that RATrain achieves up to 1.35$\times$ end-to-end speedup over MT-3000-adapted GPU-style training strategies. For LLaMA-2-7B, RATrain scales to 1024 compute clusters, reaches 112,790.55 tokens/s, and achieves 97.0\% scaling efficiency. A further 1.028B-token correctness run shows that RATrain preserves the loss trajectory of a semantically equivalent Baseline-1F1B run, with a maximum relative loss deviation of 0.081\%.
\end{abstract}


\maketitle

\section{Introduction}

Supercomputing infrastructures are evolving from platforms primarily designed for numerical simulation into integrated infrastructures that support both scientific computing and intelligent computing~\cite{jumper2021alphafold,bi2023pangu,lam2023graphcast}. As AI for Science, scientific foundation models, and large language model (LLM) training move into production supercomputing environments, heterogeneous supercomputing platforms are increasingly expected to host Transformer training workloads~\cite{shoeybi2019megatron,narayanan2021megatron}. MT-3000 is a representative heterogeneous processor used in exascale supercomputing platforms. Its compute clusters provide high FP16 compute capability, but are also constrained by an explicit memory hierarchy, limited local memory capacity, and limited inter-cluster communication bandwidth. This raises a fundamental systems question: can LLM training runtimes designed for GPU clusters still run effectively on such bandwidth-constrained heterogeneous supercomputing platforms?

Existing LLM training systems are primarily designed around GPU clusters, typically assuming high-bandwidth device memory, fast interconnects, mature collective communication libraries, and highly optimized software stacks~\cite{shoeybi2019megatron,narayanan2021megatron,rasley2020deepspeed}. On bandwidth-constrained heterogeneous platforms with explicit memory management, such as MT-3000, these assumptions are difficult to satisfy directly. GPU-oriented strategies such as TP-heavy execution, ZeRO-style sharding, and activation checkpointing can expose intra-layer communication, parameter-view reconstruction, and backward-time recovery overheads, respectively~\cite{rajbhandari2020zero,huang2019gpipe,narayanan2019pipedream,chen2016checkpoint,checkmate2020,capuchin2020,yuan2024activation,huang2025obscura}. Meanwhile, existing GEMM optimizations on multi-core DSPs mainly target FP32 general GEMM or single-operator scenarios~\cite{yu2024optimizing}, and cannot directly provide efficient FP16 GEMM, Attention Backward, and explicit data movement support for the LLM training critical path. Therefore, the bottleneck of dense LLM training on the target platform comes from the coupling among critical-path execution, parallel organization, training-state lifecycles, and platform resource constraints.
The key observation of this paper is that, on MT-3000, the asymmetry between forward and backward execution time creates stage-local scheduling windows in the standard non-interleaved 1F1B pipeline. At the same time, training states have deterministic lifecycle dependencies induced by the layer-wise structure of Transformers: gradients are produced following the backward layer order, updated parameter views are consumed by the next forward pass in layer order, and activation recovery can be completed in local windows before the corresponding backward computation arrives. Thus, dense LLM training should be modeled as training-state lifecycle scheduling, rather than coarse-grained step-end state processing.

Based on this observation, this paper presents RATrain, a resource-aware training runtime for dense LLMs on bandwidth-constrained heterogeneous supercomputing platforms. RATrain preserves the standard non-interleaved 1F1B main schedule and manages training states as runtime objects with explicit lifecycles. RATrain uses a layer-wise training-state pipeline to decompose bulk state processing at the accumulation boundary into per-layer state tasks; next-iteration update--prefetch scheduling to prepare updated parameter views according to the next forward access order; and forward-side activation recovery to move part of activation recovery from the backward critical path into local available windows. RATrain further integrates an MT-3000-aware execution backend with a resource-aware configuration planner: the former provides efficient and predictable FP16 GEMM, Attention Backward, and explicit data movement implementations for the training critical path, while the latter selects resource-feasible training configurations under the per-cluster usable-DDR constraint.

We implement and evaluate RATrain on a real MT-3000 heterogeneous supercomputing platform, using dense decoder-only configurations corresponding to LLaMA-2-7B, Baichuan2-13B, Qwen2.5-32B, and LLaMA-2-70B. Results show that RATrain can execute resource-feasible dense LLM training configurations under the 20GB usable-DDR constraint per cluster, and achieves up to 1.35$\times$ end-to-end speedup over MT-3000-adapted GPU-style training strategies. For LLaMA-2-7B, RATrain scales from 256 to 1024 compute clusters, reaching 112,790.55 tokens/s with 97.0\% scaling efficiency. A further 1.028B-token correctness run shows that RATrain preserves the loss trajectory of a semantically equivalent Baseline-1F1B run, with a maximum relative loss deviation of 0.081\%.

This paper makes the following contributions:
\begin{itemize}
\item We identify a systematic mismatch between GPU-oriented LLM training runtimes and bandwidth-constrained heterogeneous supercomputing platforms, showing that dense LLM training is bottlenecked by the coupling among critical-path execution, parallel execution, training-state lifecycles, and platform resource constraints.

\item We propose training-state lifecycle scheduling, which models gradient synchronization, parameter update, parameter-view prefetching, and activation recovery in standard non-interleaved 1F1B training as layer-level and stage-local schedulable runtime tasks.
\item We design and implement RATrain, which unifies stage-local state scheduling, an efficient training-critical-path backend, and resource-aware configuration planning in a single runtime framework to support resource-feasible dense LLM training under the per-cluster usable-DDR constraint.
\item We evaluate RATrain on a real MT-3000 platform and demonstrate its benefits in end-to-end performance, scalability, mechanism effectiveness, resource feasibility, and training-semantics preservation.
\end{itemize}

\section{Background and Motivation}
\subsection{LLM Training Constraints on MT-3000 Acceleration Clusters}
\begin{figure}[t]
  \centering
  \includegraphics[width=\linewidth]{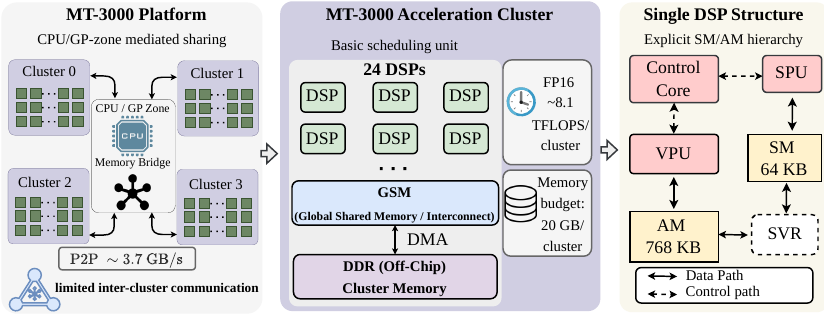}
  \caption{MT-3000 platform organization.
The platform consists of autonomous acceleration clusters connected through the CPU/GP Zone; each cluster contains 24 DSPs with an explicit SM/AM--GSM--DDR memory hierarchy.
These constraints motivate resource-aware training-state scheduling.}
  \label{fig:mt3000-platform}
\end{figure}
This paper targets MT-3000-based bandwidth-constrained heterogeneous supercomputing platforms, and uses the acceleration cluster as the basic unit for resource modeling and runtime scheduling. As shown in Fig.~\ref{fig:mt3000-platform}, an MT-3000 platform contains multiple autonomous acceleration clusters, where inter-cluster data sharing is coordinated through the CPU/GP Zone and memory bridge. Our runtime-level profiling shows that the MPI point-to-point bandwidth between clusters is about 3.7GB/s, indicating that inter-cluster communication is a constrained resource that must be explicitly modeled for dense LLM training.

A single MT-3000 acceleration cluster contains 24 DSPs. DSPs within a cluster share data through GSM and exchange data with off-chip DDR through DMA; each DSP further provides a 64KB scalar memory (SM) and a 768KB array memory (AM). Hardware profiling shows that an acceleration cluster has a theoretical FP16 peak of about 8.1 TFLOPS at 1.8GHz, but only a 20GB training memory budget and an effective DDR bandwidth of about 30GB/s. These characteristics indicate that dense LLM training on MT-3000 is not limited only by operator compute throughput, but also by local state residency, DDR data movement, and exposed inter-cluster communication.

\subsection{Training-State Lifecycles in Dense LLM Training}
\begin{figure*}[t]
  \centering
  \includegraphics[width=0.9\linewidth]{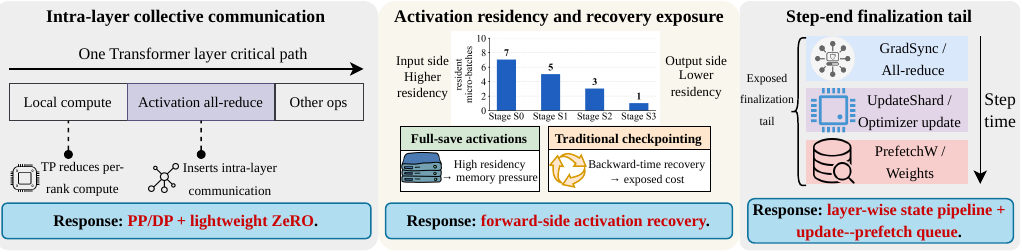}
  \caption{Motivation for training-state lifecycle scheduling on MT-3000.
Bandwidth and memory constraints expose intra-layer communication, activation recovery, and step-end state-processing costs.
RATrain mitigates them through resource-aware parallelization and layer-wise state scheduling.}
  \label{fig:motivation}
\end{figure*}
Dense decoder-only LLMs consist of sequentially stacked Transformer layers, whose forward and backward passes follow stable and opposite layer-wise access orders~\cite{vaswani2017attention,brown2020language,touvron2023llama2}. This structure gives training states naturally layered lifecycles: the residency time, access order, and dependencies of different states are jointly determined by the layer order and the 1F1B pipeline, creating exploitable stage-local scheduling windows.

On GPU clusters, high device-memory bandwidth, large device-memory capacity, and fast interconnects can partially absorb state access and communication overheads. On MT-3000, however, limited local memory, limited DDR bandwidth, and constrained inter-cluster communication amplify state residency, data movement, and synchronization exposure. Therefore, dense LLM training should not be treated merely as an operator acceleration problem, but as a training-state lifecycle scheduling problem.

\subsection{Limitations of GPU-Oriented Training Strategies}

Directly porting a GPU-oriented training runtime to MT-3000 does not provide robust benefits. Tensor parallelism scales training by introducing intra-layer collectives~\cite{shoeybi2019megatron,narayanan2021megatron}, ZeRO-style state partitioning reduces memory redundancy through parameter-view reconstruction~\cite{rajbhandari2020zero,rasley2020deepspeed}, and activation checkpointing trades activation memory for backward-time recomputation~\cite{chen2016checkpoint,checkmate2020,yuan2024activation,huang2025obscura}. On bandwidth-constrained heterogeneous platforms, these costs are more likely to be exposed on the critical path.

Fig.~\ref{fig:motivation} summarizes three representative mismatches. First, TP-heavy execution inserts intra-layer collective communication into the critical path of each Transformer layer. Second, the standard non-interleaved 1F1B pipeline creates activation residency imbalance: full-save activations increase the memory pressure on input-side stages, while backward-time recovery exposes recomputation overhead. Third, if state tasks such as GradSync, UpdateShard, and PrefetchW are delayed to the accumulation boundary, a visible finalization tail appears at the end of each step.

These observations show that the bottleneck on the target platform is not a single GEMM, Attention, or communication kernel, but the coupling among parallel execution, training-state lifecycles, and platform resource constraints. RATrain therefore formulates dense LLM training as a training-state lifecycle scheduling problem under the standard non-interleaved 1F1B pipeline. Together with an MT-3000-aware operator backend, RATrain reduces exposed overhead through layer-level and stage-local state scheduling.

\section{RATrain Design Overview}

\subsection{Design Overview}

RATrain is a resource-aware training runtime for dense LLMs on bandwidth-constrained heterogeneous supercomputing platforms. It preserves the standard non-interleaved 1F1B execution order, but does not treat training-state processing as centralized step-end work. Instead, it schedules training-state lifecycles around the layer order and stage-local windows. The key design choice of RATrain is to make state operations locally schedulable while preserving training semantics. By moving communication, parameter preparation, and activation recovery into overlappable windows, RATrain reduces their exposure at the step boundary or on the backward critical path.

\subsection{System Architecture}
\begin{figure}[t]
  \centering
  \includegraphics[width=\linewidth]{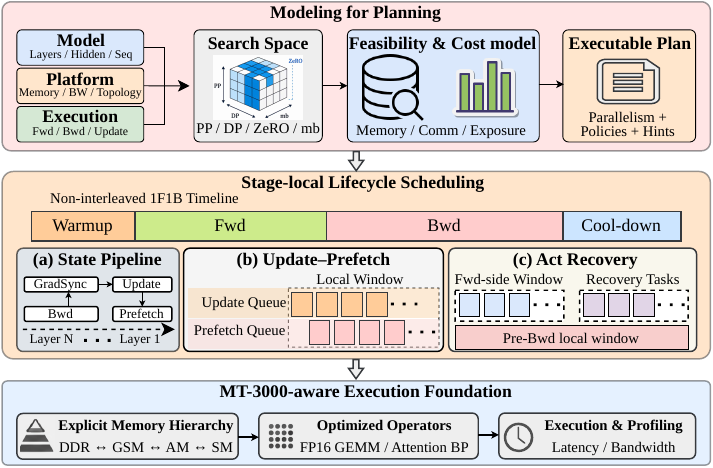}
  \caption{RATrain overview: profile-guided planning, stage-local lifecycle scheduling, and MT-3000-aware backend execution.}
  \label{fig:ratrain-overview}
\end{figure}
Fig.~\ref{fig:ratrain-overview} shows the overall architecture of RATrain. The system mainline consists of three stages. First, RATrain builds unified profiles from model structure, platform resources, and execution costs, and uses the planner to select a resource-feasible training plan. Second, the stage-local runtime performs lifecycle scheduling while preserving the standard 1F1B main path. Finally, the MT-3000 backend maps scheduled tasks to platform-aware operators, explicit data movement, and local memory management.

This architecture connects configuration selection, runtime scheduling, and platform execution. The planner defines the resource boundary of the training plan, the runtime determines when state tasks are issued within local windows, and the backend provides predictable operator and data-movement support. Together, these components allow RATrain to coordinate model scale, memory capacity, communication bandwidth, and the 1F1B timing structure.

\subsection{Execution Abstraction}
RATrain targets dense decoder-only LLM training. It partitions the model into consecutive pipeline stages along the layer dimension. Each stage is mapped to one or more MT-3000 acceleration clusters and manages the forward pass, backward pass, state tasks, and memory resources for its local layers. Different data-parallel replicas use the same stage partitioning.
During execution, RATrain preserves the standard non-interleaved 1F1B order: the forward pass propagates from the input-side stage to the output-side stage, and the backward pass returns in the opposite direction. Meanwhile, the stage-local scheduler triggers lifecycle-related state tasks only after local dependencies are satisfied. For example, state processing for a layer becomes schedulable only after local gradient accumulation for that layer completes; the corresponding parameter view must be ready before the next forward pass accesses the layer; and for a micro-batch whose backward pass is about to reach the current stage, the runtime can recover the required intermediate activations in the previous available stage-local window.

Therefore, RATrain decouples training semantics from state scheduling: the main training path remains unchanged, while state tasks are scheduled within layer-level and stage-local control spaces. The next section describes the concrete mechanisms.

\section{Design}
This section presents the concrete mechanisms of RATrain. Following the system mainline in Fig.~\ref{fig:ratrain-overview}, 
Section~\ref{sec:backend} introduces the MT-3000-aware backend that provides stable execution-latency profiles for the planner and runtime; 
Section~\ref{sec:state-pipeline} presents the layer-wise state pipeline and update--prefetch scheduling; 
Section~\ref{sec:activation-recovery} describes FSR; 
and Section~\ref{sec:planner} introduces the resource-aware configuration planner.
\subsection{MT-3000-aware FP16 GEMM and Attention Backward Backend}
\label{sec:backend}

RATrain's lifecycle scheduling relies on stable and predictable layer-level execution latency. On MT-3000, the cost of a Transformer layer is not determined only by MAC computation, but is also affected by explicit data movement among DDR, GSM, AM, and SM. Therefore, RATrain implements an MT-3000-aware operator backend beneath the runtime, providing stable latency profiles for the planner and the stage-local runtime. This backend mainly covers two training-critical paths: the FP16 GEMM assembly pipeline and memory-resident Attention Backward. The former provides stable primitives for linear layers, FFN layers, and internal matrix multiplications in Attention BP, while the latter reorganizes Attention BP into a tile schedule aware of the explicit memory hierarchy to reduce execution latency on the backward path.
\begin{figure}[t]
  \centering
  \includegraphics[width=\linewidth]{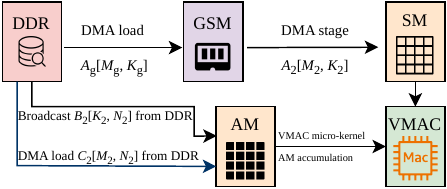}
  \caption{FP16 GEMM dataflow on MT-3000. RATrain stages $A$ through GSM/SM, broadcasts $B$ to AM, and accumulates $C$ in AM during VMAC execution.}
  \label{fig:gemm-backend}
\end{figure}

\textbf{FP16 GEMM assembly pipeline.}
The QKV projection, output projection, FFN projection, and internal matrix multiplications in Attention BP can all be reduced to GEMM primitives. RATrain decomposes the GEMM backend into cluster-level dataflow and a DSP-local assembly pipeline. The former organizes tile movement and reuse around DDR, GSM, AM, and SM, while the latter uses VLIW instruction-level scheduling to reduce functional-unit conflicts. Fig.~\ref{fig:gemm-backend} shows the GEMM dataflow of RATrain on MT-3000. RATrain first loads $A_g[M_g,K_g]$ from DDR into GSM, and then stages $A_2[M_2,K_2]$ into SM. Meanwhile, $B_2[K_2,N_2]$ is broadcast from DDR to AM, and the output tile $C_2[M_2,N_2]$ is loaded from DDR into AM as the accumulation buffer. The VMAC micro-kernel then consumes operands from SM and AM to perform FP16 MAC, accumulates the result in AM, and finally writes it back to DDR through DMA.
\begin{algorithm}[t]
\small
\caption{Memory-resident Attention Backward Tile Schedule}
\label{alg:attention-bp}
\begin{algorithmic}[1]
\Require Query tile $Q_i$, output gradient $GO_i$, saved probability tiles $\{P_{ij}\}$, key/value tiles $\{K_j,V_j\}$
\Ensure Query gradient $GQ_i$, key/value gradients $\{GV_j,GK_j\}$

\State \textbf{Outer-resident setup:} $(Q_i, GO_i) \gets \Call{LoadAM}{Q_i, GO_i}$
\State \textbf{Allocate} AM buffer for $GQ_i$ and initialize it to zero

\For{each key/value tile $j$}
    \State \textbf{Inner-loop broadcast:} $(K_j,V_j) \gets \Call{BcastAM}{K_j,V_j}$
    \State \textbf{Forward-state load:} $P_{ij} \gets \Call{LoadAM}{P_{ij}}$

    \State \textbf{AM-resident compute:} 
    $GP_{ij} \gets GO_i V_j^{T}$,
    $GS_{ij} \gets \mathrm{SoftmaxBackward}(P_{ij}, GP_{ij})$

    \State \textbf{SM staging for $GV_j$:}
    $\widetilde{P}_{ij}^{T} \gets \Call{StageSM}{P_{ij}^{T}}$
    \State $GV_j^{\mathrm{part}} \gets \widetilde{P}_{ij}^{T}GO_i$
    \State \textbf{GSM reduction:} 
    $GV_j \gets \Call{ReduceAddGSM}{GV_j, GV_j^{\mathrm{part}}}$

    \State \textbf{SM staging for $GQ_i$:}
    $\widetilde{GS}_{ij} \gets \Call{StageSM}{GS_{ij}}$
    \State $GQ_i \gets GQ_i + \widetilde{GS}_{ij}K_j$

    \State \textbf{SM staging for $GK_j$:}
    $\widetilde{GS}_{ij}^{T} \gets \Call{StageSM}{GS_{ij}^{T}}$
    \State $GK_j^{\mathrm{part}} \gets \widetilde{GS}_{ij}^{T}Q_i$
    \State \textbf{GSM reduction:}
    $GK_j \gets \Call{ReduceAddGSM}{GK_j, GK_j^{\mathrm{part}}}$
\EndFor

\State \textbf{Writeback:} \Call{WriteBack}{$GQ_i$}
\end{algorithmic}
\end{algorithm}
In the DSP-local assembly pipeline, RATrain interleaves address generation, load, half-precision extraction, broadcast, and FP16 MAC, so that the preparation of the next $A_{\mathrm{next}}/B_{\mathrm{next}}$ operands overlaps with the current MAC. This reduces VLIW functional-unit conflicts and memory-access bubbles. Table~\ref{tab:gemm-pipeline} gives the complete micro-kernel pipeline.

\begin{table*}[t]
\centering
\scriptsize
\setlength{\tabcolsep}{2.5pt}
\renewcommand{\arraystretch}{0.92}
\caption{Complete assembly pipeline organization of the GEMM micro-kernel.}
\label{tab:gemm-pipeline}
\begin{tabular}{llllll}
\toprule
\textbf{VMAC} & \textbf{SMAC1/2} & \textbf{SLDST} & \textbf{VLDST1/2} & \textbf{SIEU} & \textbf{SBR} \\
\midrule
\texttt{vfmulas32 A[0,1,2][0], B[0][0]} & \texttt{smvaga A\_next} &  & \texttt{vldw B[1][2,3]} &  &  \\
\texttt{vfmulas32 A[0,1,2][0], B[0][1]} &  &  &  & \texttt{sbale A[0][1]} &  \\
\texttt{vfmulas32 A[0,1,2][0], B[0][2]} & \texttt{svbcast A[0][1]} & \texttt{sldh A\_next[0][0]} &  & \texttt{sbale A[1][1]} &  \\
\texttt{vfmulas32 A[0,1,2][0], B[0][3]} & \texttt{svbcast A[1][1]} & \texttt{sldh A\_next[1][0]} &  & \texttt{sbale A[2][1]} &  \\
\texttt{vfmulas32 A[3,4,5][0], B[0][0]} & \texttt{svbcast A[2][1] | sadd B\_next} & \texttt{sldh A\_next[2][0]} &  & \texttt{sbale A[3][1]} &  \\
\texttt{vfmulas32 A[3,4,5][0], B[0][1]} & \texttt{svbcast A[3][1] | smvaga B\_next} & \texttt{sldh A\_next[3][0]} &  & \texttt{sbale A[4][1]} &  \\
\texttt{vfmulas32 A[3,4,5][0], B[0][2]} & \texttt{svbcast A[4][1]} & \texttt{sldh A\_next[4][0]} &  & \texttt{sbale A[5][1]} &  \\
\texttt{vfmulas32 A[3,4,5][0], B[0][3]} & \texttt{svbcast A[5][1]} & \texttt{sldh A\_next[5][0]} & \texttt{vldw B\_next[0][0,1]} &  &  \\
\texttt{vfmulas32 A[0,1,2][1], B[1][0]} &  &  & \texttt{vldw B\_next[0][2,3]} & \texttt{seq} &  \\
\texttt{vfmulas32 A[0,1,2][1], B[1][1]} &  &  &  & \texttt{sbale A\_next[0][0]} & \texttt{sbr} \\
\texttt{vfmulas32 A[0,1,2][1], B[1][2]} & \texttt{svbcast A\_next[0][0]} & \texttt{sldh A\_next[0][1]} &  & \texttt{sbale A\_next[1][0]} &  \\
\texttt{vfmulas32 A[0,1,2][1], B[1][3]} & \texttt{svbcast A\_next[1][0]} & \texttt{sldh A\_next[1][1]} &  & \texttt{sbale A\_next[2][0]} &  \\
\texttt{vfmulas32 A[3,4,5][1], B[1][0]} & \texttt{svbcast A\_next[2][0]} & \texttt{sldh A\_next[2][1]} &  & \texttt{sbale A\_next[3][0]} &  \\
\texttt{vfmulas32 A[3,4,5][1], B[1][1]} & \texttt{svbcast A\_next[3][0]} & \texttt{sldh A\_next[3][1]} &  & \texttt{sbale A\_next[4][0]} &  \\
\texttt{vfmulas32 A[3,4,5][1], B[1][2]} & \texttt{svbcast A\_next[4][0]} & \texttt{sldh A\_next[4][1]} &  & \texttt{sbale A\_next[5][0]} &  \\
\texttt{vfmulas32 A[3,4,5][1], B[1][3]} & \texttt{svbcast A\_next[5][0] | sadd A\_next} & \texttt{sldh A\_next[5][1]} & \texttt{vldw B\_next[1][0,1]} &  &  \\
\bottomrule
\end{tabular}
\end{table*}

\textbf{Memory-resident Attention Backward.}
Attention BP includes the computation of $dV$, $dP$, $dS$, $dQ$, and $dK$, tile transposition, softmax backward, and cross-DSP gradient accumulation. If these operations are split into multiple independent kernels and intermediate states such as $P_{ij}$, $P_{ij}^{T}$, $GP_{ij}$, $GS_{ij}$, and $GS_{ij}^{T}$ are passed through DDR, the backward path incurs substantial off-chip traffic and extra latency. RATrain therefore organizes Attention BP as a memory-resident tile schedule, as shown in Algorithm~\ref{alg:attention-bp}.

This schedule adopts a query-outer and K/V-inner loop structure. For an outer query tile, $Q_i$, $GO_i$, and necessary forward states are kept resident in AM as much as possible during the inner K/V loop, reducing repeated query-side movement. For an inner K/V tile, $K_j/V_j$ is broadcast from DDR to 24 AMs and consumed by multiple DSPs. RATrain directly loads the probability tile $P_{ij}$ saved during the forward pass, instead of recomputing it on the backward critical path. In tile-local computation, $P_{ij}^{T}$, $GS_{ij}$, and $GS_{ij}^{T}$ are micro-tiled and staged into SM as GEMM left operands. The partial sums of $GV_j$ and $GK_j$ are then reduced across 24 DSPs through GSM and written back to DDR.

RATrain uses explicit capacity constraints to select executable tile configurations. For a candidate tile shape, the AM working set, SM micro-staging buffer, and GSM reduction block must satisfy:
\begin{equation}
\begin{aligned}
D_{\mathrm{size}}(B_rB_c + 2B_rd + 2B_cd) &\le C_{\mathrm{AM}}, \\
D_{\mathrm{size}}B_c'B_r &\le C_{\mathrm{SM}}, \\
D_{\mathrm{size}}B_r'B_c &\le C_{\mathrm{SM}}, \\
D_{\mathrm{size}}B_cd &\le C_{\mathrm{GSM}}.
\end{aligned}
\label{eq:attn-bp-capacity}
\end{equation}
These constraints bound the attention tile and resident operands in AM, the micro-staged left operands in SM, and the cross-DSP reduction block in GSM, respectively. This schedule differs from FlashAttention, which primarily optimizes GPU HBM access. RATrain instead targets the MT-3000 backward path, where the key issues are K/V broadcast reuse, SM-limited left-operand staging, and GSM-based local gradient reduction.
\subsection{Layer-wise State Pipeline and Update--Prefetch Scheduling}
\label{sec:state-pipeline}

In conventional data-parallel or ZeRO-style training, gradient synchronization, parameter update, and preparation of the next-round parameter view are usually deferred to the accumulation boundary, forming a bulk state-processing phase at the end of a step. RATrain exploits the layer-wise order of Transformer backward execution to decompose these state operations into layer-level, stage-local lifecycle tasks, and schedules them according to the access order of the next forward pass.
\begin{figure}[t]
  \centering
  \includegraphics[width=\linewidth]{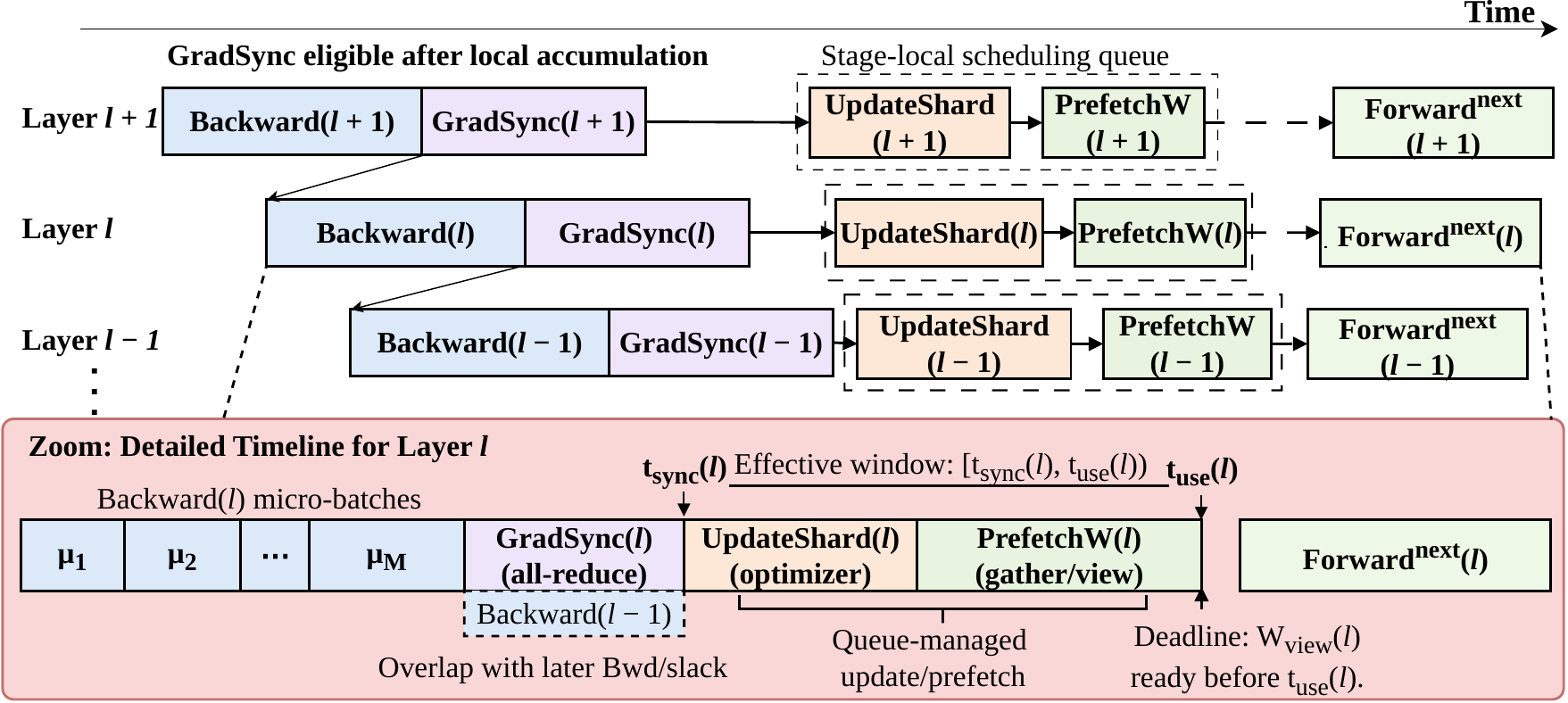}
  \caption{Layer-wise state pipeline and update--prefetch scheduling. $\mathrm{GradSync}$ overlaps with later backward/slack, while $\mathrm{UpdateShard}$ and $\mathrm{PrefetchW}$ are queue-managed to prepare $W_{\mathrm{view}}$ before the next forward access.}
  \label{fig:state-pipeline}
\end{figure}
Fig.~\ref{fig:state-pipeline} shows the layer-wise state pipeline in RATrain. For layer $l$, $\mathrm{GradSync}(l)$ becomes schedulable only after the local gradient accumulation of this layer completes within the current accumulation window. It is not triggered immediately after a single micro-batch $\mathrm{Backward}(l)$. The stage-local scheduler then tries to overlap $\mathrm{GradSync}(l)$ with subsequent backward computation or stage-local slack, reducing the finalization tail at the end of the step.

After $\mathrm{GradSync}(l)$ completes, the layer enters the following state-task chain:
\begin{equation}
\mathrm{UpdateShard}(l)
\rightarrow
\mathrm{PrefetchW}(l).
\label{eq:state-task-chain}
\end{equation}

Here, $\mathrm{UpdateShard}(l)$ updates the parameter shard and optimizer states of the layer, while $\mathrm{PrefetchW}(l)$ prepares the updated working weight view for the next forward pass. This task chain only changes where state tasks are scheduled after their dependencies are satisfied; it does not change the forward/backward order, gradient accumulation rule, or optimizer update semantics.
To avoid stalling the next forward pass on parameter-view preparation, RATrain treats update--prefetch as a deadline-aware scheduling problem. Let $t_{\mathrm{sync}}(l)$ denote the completion time of $\mathrm{GradSync}(l)$, and let $t_{\mathrm{use}}(l)$ denote the expected time when the next $\mathrm{Forward}(l)$ accesses $W_{\mathrm{view}}(l)$. The effective scheduling window is:
\begin{equation}
t_{\mathrm{sync}}(l) \le t < t_{\mathrm{use}}(l).
\label{eq:update-prefetch-window}
\end{equation}

Within this window, RATrain schedules $\mathrm{UpdateShard}(l)$ and $\mathrm{PrefetchW}(l)$ in order. If $\mathrm{PrefetchW}(l)$ finishes before $t_{\mathrm{use}}(l)$, the next forward pass can directly use $W_{\mathrm{view}}(l)$; otherwise, the uncovered portion appears as a next-forward stall. The exposed latency is estimated as:
\begin{equation}
\begin{aligned}
E_{\mathrm{upd}}(l)
&=
\max\bigl(0, T_{\mathrm{upd}}(l)-W_{\mathrm{upd}}(l)\bigr),\\
E_{\mathrm{pref}}(l)
&=
\max\bigl(0, T_{\mathrm{pref}}(l)-W_{\mathrm{pref}}(l)\bigr).
\end{aligned}
\label{eq:update-prefetch-exposure}
\end{equation}

Here, $T_{\mathrm{upd}}$ and $T_{\mathrm{pref}}$ denote the update and prefetch latency, while $W_{\mathrm{upd}}$ and $W_{\mathrm{pref}}$ denote the stage-local windows available for hiding them. In this way, RATrain converts bulk state processing at the accumulation boundary into layer-wise lifecycle scheduling, reducing the finalization tail.

\subsection{Forward-Side Activation Recovery}
\label{sec:activation-recovery}

Besides parameter and gradient states, activations are also a major source of memory consumption in 1F1B training. In non-interleaved 1F1B, input-side stages usually need to keep forward activations for more in-flight micro-batches. A full-save policy avoids recovery overhead but incurs high peak memory usage. Conventional checkpointing reduces the amount of resident activations, but missing intermediate states are typically recovered only after backward reaches the current stage, exposing the recovery latency on the backward critical path.

RATrain proposes FSR. The key idea is to move activation recovery to an available window before backward arrives, while preserving the standard 1F1B forward/backward order. For layers using recovery, RATrain keeps only necessary checkpoints after forward and releases recoverable temporary activations. When the backward of a micro-batch is about to return to the current stage, the runtime recovers the intermediate states required by backward in a previous available forward-side or idle slot. As a result, when the current stage starts the corresponding backward computation, the required activations are already ready, and the recovery latency is no longer fully added to the backward critical path.
\begin{figure}[t]
  \centering
  \includegraphics[width=\linewidth]{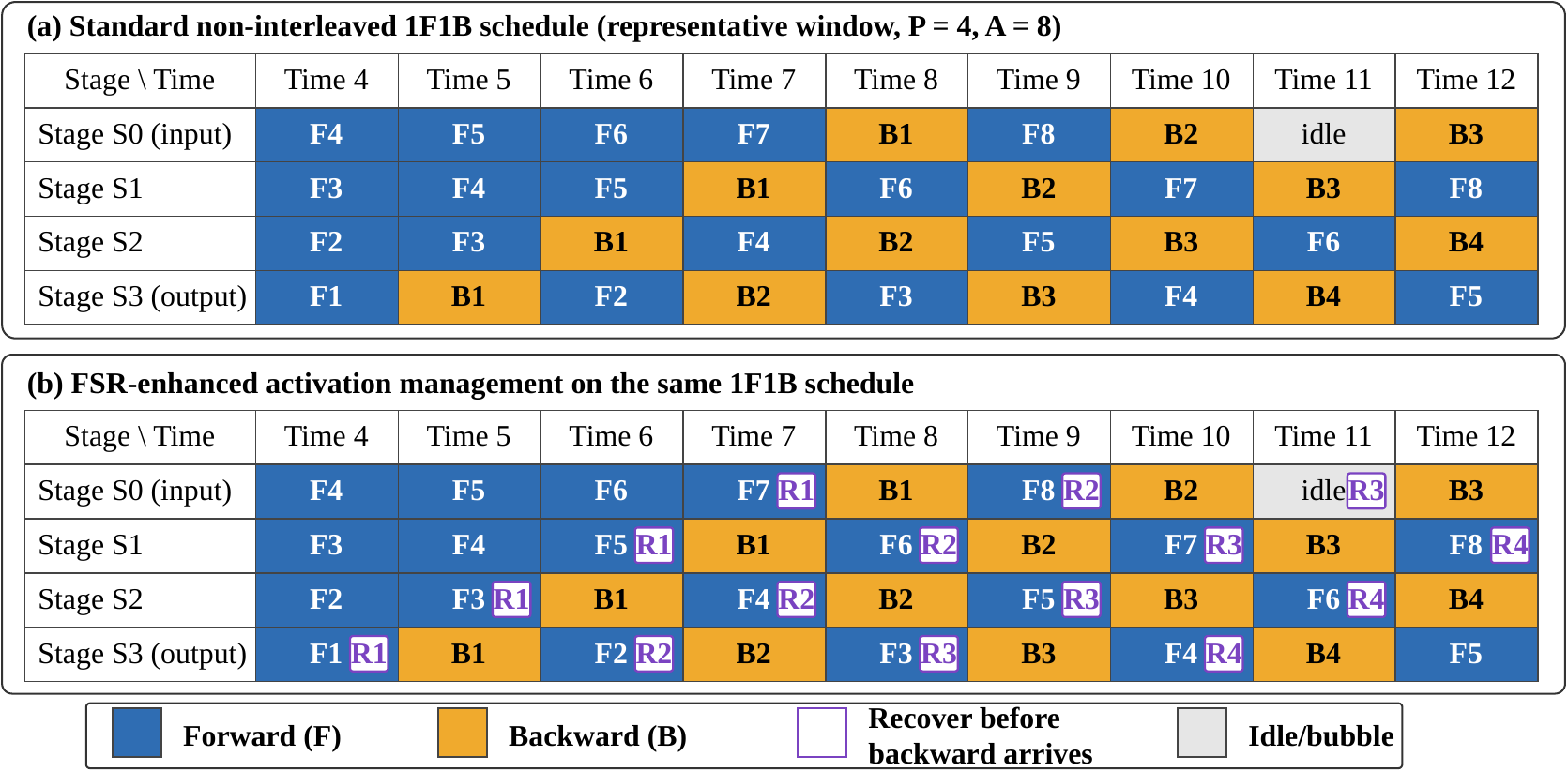}
  \caption{FSR on standard non-interleaved 1F1B. RATrain keeps only checkpoints after forward and recovers missing activations in a forward-side or idle slot before the corresponding backward reaches the stage.}
  \label{fig:fsr}
\end{figure}
Fig.~\ref{fig:fsr} shows FSR on the standard non-interleaved 1F1B schedule. RATrain does not change the main 1F1B execution order. Instead, it changes the placement of recovery tasks: missing activations are recovered before backward arrives and are delivered to the subsequent backward computation through a short-lived recovery buffer.

Let $N_{\mathrm{act}}(p)$ denote the number of micro-batches whose activations must be resident at stage $p$ under 1F1B. Let $M_{\mathrm{full}}$ be the full activation size of one micro-batch, $M_{\mathrm{ckpt}}$ be the checkpoint size, and $M_{\mathrm{rec}}$ be the recovery-buffer size. The activation peak of full-save can be approximated as:
\begin{equation}
M_{\mathrm{act,full}}(p)
\approx
N_{\mathrm{act}}(p) M_{\mathrm{full}} .
\label{eq:act-full}
\end{equation}

With FSR, long-lived full activations are replaced by multiple checkpoints and one short-lived recovery buffer. The peak memory is approximated as:
\begin{equation}
M_{\mathrm{act,FSR}}(p)
\approx
N_{\mathrm{act}}(p) M_{\mathrm{ckpt}} + M_{\mathrm{rec}} .
\label{eq:act-fsr}
\end{equation}

Since $M_{\mathrm{ckpt}}$ is usually much smaller than $M_{\mathrm{full}}$, FSR reduces the activation peak of input-side stages. RATrain does not assume that recovery can always be fully hidden. When the forward-side recovery window is insufficient or local resources are unavailable, the runtime can fall back to backward-time recovery. This fallback preserves training semantics, while the uncovered recovery latency is included in the step-time estimate:
\begin{equation}
E_{\mathrm{rec}}(p)
=
\max\bigl(0, T_{\mathrm{rec}}(p)-W_{\mathrm{rec}}(p)\bigr).
\label{eq:recovery-exposure}
\end{equation}

Here, $T_{\mathrm{rec}}(p)$ denotes the activation recovery latency of stage $p$, and $W_{\mathrm{rec}}(p)$ denotes the stage-local window available for hiding the recovery task.

\subsection{Resource-Aware Configuration Planner}
\label{sec:planner}

RATrain's runtime mechanisms do not work in isolation; their benefits depend on the match between training configurations and platform resources. Different parallelization choices and runtime policies jointly affect stage memory pressure, the 1F1B timing structure, and the windows in which state tasks can be hidden. Fixed heuristics are therefore difficult to apply robustly across different model sizes and resource constraints. RATrain uses a resource-aware configuration planner to filter memory-feasible training plans from the candidate space and select the configuration with the lowest estimated step time.

RATrain represents a candidate training configuration as:
\begin{equation}
c = (P, D, Z, b, A, \pi_{\mathrm{act}}, \pi_{\mathrm{pref}}),
\label{eq:config}
\end{equation}
where $P$ is the pipeline degree, $D$ is the data-parallel degree, $Z$ is the ZeRO stage, $b$ is the local micro-batch size, $A$ is the number of gradient accumulation steps, and $\pi_{\mathrm{act}}$ and $\pi_{\mathrm{pref}}$ denote the activation recovery policy and parameter prefetch policy, respectively. The planner takes the model profile, platform profile, and execution profile as input, and searches the candidate space $\mathcal{C}$ for a configuration that satisfies the resource constraints and minimizes the estimated step time.

The planner first performs memory-feasibility pruning. For a candidate configuration $c$, the peak memory of stage $p$ is estimated as:
\begin{equation}
M_p(c)
=
M_{\mathrm{state}}^p(c)
+
M_{\mathrm{act}}^p(c,\pi_{\mathrm{act}})
+
M_{\mathrm{buf}}^p(c,\pi_{\mathrm{pref}},\pi_{\mathrm{act}}).
\label{eq:planner-memory}
\end{equation}
Here, $M_{\mathrm{state}}^p(c)$ includes local parameter shards, gradient states, and optimizer states; $M_{\mathrm{act}}^p(c,\pi_{\mathrm{act}})$ captures the activation residency under full-save, checkpointing, or FSR; and $M_{\mathrm{buf}}^p(c,\pi_{\mathrm{pref}},\pi_{\mathrm{act}})$ includes short-lived buffers for communication, prefetching, recovery, and operator execution. A candidate is feasible only if:
\begin{equation}
\max_p M_p(c) \le M_{\mathrm{budget}}.
\label{eq:planner-memory-constraint}
\end{equation}

For memory-feasible configurations, the planner estimates the step time. RATrain uses an exposed-latency decomposition to capture how stage-local tasks affect the end-to-end step time. For any schedulable task $x$, its exposed latency is defined as:
\begin{equation}
E_x(c) = \max\bigl(0, T_x(c)-W_x(c)\bigr),
\label{eq:exposed-latency}
\end{equation}
where $T_x(c)$ is the task latency from the execution profile, and $W_x(c)$ is the time that can be covered by the 1F1B timing structure, stage-local slack, or bounded scheduling windows. If a task is fully covered by computation overlap, communication overlap, prefetch windows, or recovery windows, its exposed latency is zero; otherwise, the uncovered portion contributes to the step time.

Based on this definition, the step time of a candidate configuration is estimated as:
\begin{equation}
T_{\mathrm{step}}(c)
=
T_{\mathrm{1F1B}}(c)
+
E_{\mathrm{comm}}(c)
+
E_{\mathrm{upd}}(c)
+
E_{\mathrm{pref}}(c)
+
E_{\mathrm{rec}}(c).
\label{eq:planner-step-time}
\end{equation}
Here, $T_{\mathrm{1F1B}}(c)$ denotes the main execution time of standard non-interleaved 1F1B, including forward/backward slot time, pipeline bubbles, and stage imbalance. $E_{\mathrm{comm}}(c)$ captures exposed communication from stage-boundary transfers and GradSync. Under this decomposition, the residual tail near the accumulation boundary is represented by the uncovered communication, update, and prefetch costs, rather than being counted as a separate term.

Finally, the planner solves the following constrained selection problem:
\begin{equation}
c^\star =
\arg\min_{c\in\mathcal{C}} T_{\mathrm{step}}(c),
\quad
\mathrm{s.t.}\quad
\max_p M_p(c) \le M_{\mathrm{budget}}.
\label{eq:planner-objective}
\end{equation}

Algorithm~\ref{alg:planner} summarizes the planning procedure. The planner is not intended to replace end-to-end measurement. Instead, it uses profiles collected on the same platform before training to prune the large configuration space and passes the selected training plan to the runtime. The plan includes the PP/DP/ZeRO combination, stage partitioning, micro-batch and accumulation settings, activation policy, prefetch policy, and necessary runtime scheduling hints.

\begin{algorithm}[]
\small
\caption{Resource-aware Configuration Planning}
\label{alg:planner}
\begin{algorithmic}[1]
\Require Model profile, platform profile, execution profile, search space $\mathcal{C}$
\Ensure Selected training plan $c^\star$
\State $\mathcal{V} \gets \emptyset$
\For{each candidate $c \in \mathcal{C}$}
    \State Partition layers according to the pipeline degree $P$
    \State Estimate stage memory $M_p(c)$ for each stage $p$
    \If{$\max_p M_p(c) > M_{\mathrm{budget}}$}
        \State \textbf{continue}
    \EndIf
    \State Estimate $T_{\mathrm{1F1B}}(c)$ from forward/backward profiles
    \State Estimate exposed latencies $E_{\mathrm{comm}}$, $E_{\mathrm{upd}}$, $E_{\mathrm{pref}}$, and $E_{\mathrm{rec}}$
    \State $T_{\mathrm{step}}(c) \gets T_{\mathrm{1F1B}}(c)
        + E_{\mathrm{comm}}(c)
        + E_{\mathrm{upd}}(c)
        + E_{\mathrm{pref}}(c)
        + E_{\mathrm{rec}}(c)$
    \State Insert $(c,T_{\mathrm{step}}(c))$ into $\mathcal{V}$
\EndFor
\State \Return $c^\star \gets \arg\min_{(c,T)\in\mathcal{V}} T$
\end{algorithmic}
\end{algorithm}

\section{Implementation}
\label{sec:implementation}

\subsection{Stage-Local Runtime}
\label{sec:stage-local-runtime}

RATrain implements each pipeline stage as a lightweight stage-local runtime. Each runtime is bound to one or more MT-3000 acceleration clusters and executes local forward, backward, and state tasks according to the selected training plan generated by the planner. Global synchronization is kept only at necessary points, such as step initialization, stage-boundary communication, and the accumulation boundary. Layer-level state tasks and activation recovery tasks are scheduled independently by each stage according to local dependencies.

Each runtime maintains local task queues for the 1F1B main path, $\mathrm{GradSync}$, $\mathrm{UpdateShard}$, $\mathrm{PrefetchW}$, and activation recovery. Queue entries are triggered by local events, such as layer backward completion, local gradient accumulation completion, parameter update completion, next-forward access deadline, and backward-arrival deadline. This design avoids global fine-grained scheduling and matches the MT-3000 hardware organization, where the acceleration cluster is the basic execution unit.

\subsection{Memory and Communication Management}
\label{sec:memory-communication}

MT-3000 exposes an explicit memory hierarchy and limited per-cluster memory. RATrain therefore implements a memory manager in each stage-local runtime to manage data objects with different lifetimes. Long-lived objects include parameter shards, optimizer states, and metadata; medium-lived objects include checkpoints, gradient buckets, and working-weight buffers; short-lived objects include temporary activations, communication staging buffers, recovery buffers, and operator workspaces. The runtime pre-allocates major buffers according to the selected training plan and reuses short-lived memory across recovery, prefetch, and operator execution.

The communication layer mainly handles stage-boundary activation transfer, data-parallel $\mathrm{GradSync}$, and parameter-view materialization. RATrain does not assume that these transfers can proceed without resource conflicts. When the communication channel or staging buffers are contended, the runtime prioritizes stage-boundary transfers on the 1F1B critical path, while other communication tasks are scheduled only when their dependencies are satisfied and resources are available.

\subsection{Operator Integration and Semantics}
\label{sec:operator-semantics}

RATrain materializes the selected training plan into stage-local execution descriptors. Each layer descriptor records the tensor shape, parameter shard layout, gradient bucket layout, activation policy, prefetch policy, and workspace requirement. The runtime generates forward/backward tasks, $\mathrm{GradSync}$ tasks, $\mathrm{UpdateShard}$ tasks, $\mathrm{PrefetchW}$ tasks, and recovery tasks from these descriptors, and attaches them to the corresponding local queues.

Backend binding is also driven by the layer descriptor. The runtime invokes MT-3000-aware GEMM, Attention Backward, and memory movement primitives, and allocates the required workspace across DDR, GSM, AM, and SM. RATrain does not change the model computation graph, micro-batch order, gradient accumulation rule, or optimizer update formula. It only changes the materialization time, buffer residency, and dispatch order of training-state tasks.
\section{Evaluation}
\subsection{Experimental Setup}
We evaluate RATrain on a real MT-3000 heterogeneous supercomputing platform~\cite{lu2022mt}. Except for the resource-aware planner, which uses execution profiles collected on the same platform for configuration selection, all performance, memory, and correctness results are obtained from actual runs on the target platform. Unless otherwise specified, the sequence length is set to 2048 and the available training memory budget per cluster is limited to 20GB.

Our experiments cover dense decoder-only configurations corresponding to LLaMA-2-7B/13B/70B~\cite{touvron2023llama2}, Baichuan2-13B~\cite{yang2023baichuan2}, and Qwen2.5-32B~\cite{yang2024qwen25}. We use an English C4 fixed token stream~\cite{raffel2020exploring,dodge2021documenting} as the training data. For each controlled comparison, all methods use the same data order, global-batch semantics, optimizer settings, learning-rate schedule, and gradient accumulation rules.
To isolate the effect of runtime scheduling, all MT-3000 baselines use the same operator backend, communication implementation, and explicit data-movement implementation. We compare RATrain with representative GPU-style training strategies, including TP-heavy~\cite{shoeybi2019megatron,narayanan2021megatron}, ZeRO-3-heavy~\cite{rajbhandari2020zero,rasley2020deepspeed}, Backward Ckpt~\cite{chen2016checkpoint}, Full-save, and Tuned PP/DP/ZeRO~\cite{huang2019gpipe,narayanan2019pipedream,rajbhandari2020zero}. Full RATrain enables layer-wise state pipeline, next-iteration update--prefetch scheduling, and FSR.

We report tokens/s, step time, training time, peak memory, scaling efficiency, and loss correctness. All end-to-end performance comparisons are conducted under the same MT-3000 backend and hardware constraints. A800 runs are used only for correctness validation and reference-scale comparison, rather than as strict cross-hardware performance baselines.

\subsection{Correctness Validation and Reference-Scale Comparison}

RATrain reschedules the execution of training state tasks without altering the standard training semantics of dense LLMs. To validate this, we conducted a 1.028B-token correctness run and compared RATrain with a semantically equivalent Baseline-1F1B. The experiments used the LLaMA-2-7B configuration with a sequence length of 2048 and a global batch of 2048. Both methods employed the same tokenizer, initial weights, data order, optimizer settings, learning-rate schedule, and gradient accumulation semantics.
\begin{figure}[t]
  \centering
  \includegraphics[width=\linewidth]{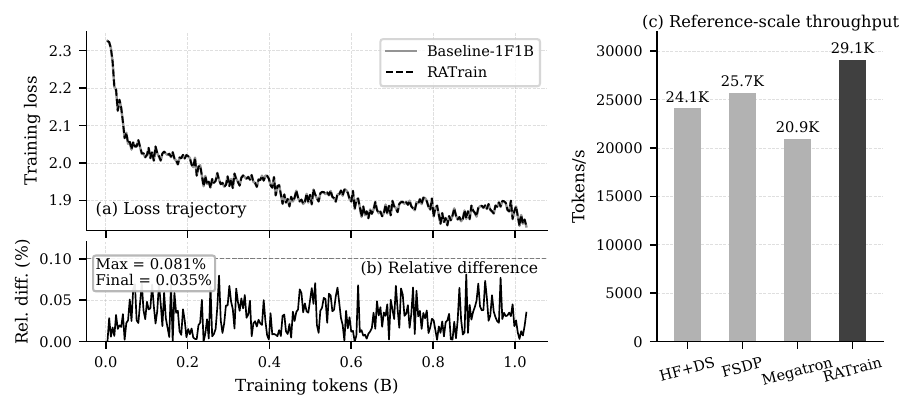}
  \caption{
  Correctness validation and reference throughput of RATrain versus Baseline-1F1B. 
  Left: training loss and per-step relative loss difference. 
  Right: throughput comparison under the same token budget; HF+DS, FSDP, and Megatron denote 8×A800 references.
  }
  \label{fig:correctness}
\end{figure}

Figure~\ref{fig:correctness} presents the training loss trajectory, per-step relative loss difference, and the reference throughput under the same token budget. RATrain and Baseline-1F1B loss curves nearly overlap entirely. The maximum, mean, and final per-step relative loss differences are 0.081\%, 0.030\%, and 0.035\%, respectively. The final losses are 1.8306 and 1.8312, with an absolute difference of only 0.00064. This result indicates that RATrain's core mechanisms change where schedulable state tasks are placed, rather than altering the computation or update semantics of dense LLM training.

To illustrate the actual scale of the correctness run, we also present reference training results on 8$\times$A800. All reference runs used the same sequence length, global batch, and 1.028B-token budget. Note that the A800 results serve only as numerical and reference-scale context, not as strict cross-hardware performance baselines. Under this setting, RATrain achieves 29,069.73 tokens/s on 256 MT-3000 compute clusters, while the three A800 reference runs reach 24,084.54, 25,702.36, and 20,914.00 tokens/s. Despite differences in architecture, memory hierarchy, and interconnect organization between MT-3000 and A800, these results indicate that MT-3000, under RATrain's resource-aware scheduling, can achieve throughput comparable to the 8$\times$A800 reference stack at the 1B-token scale. 

\subsection{End-to-End Comparison with GPU-Style Training Strategies}
\label{sec:eval-gpustyle}

This section evaluates a core system question: on the same MT-3000 backend, if the RATrain training-state lifecycle scheduling is not used and common GPU-style training strategies are adapted with tuned configurations, can comparable end-to-end training efficiency be achieved? The baselines are not direct runs of the original Megatron-LM or DeepSpeed GPU implementations, but representative strategies constructed on the MT-3000 backend with identical operator, explicit data movement, and communication implementations. Therefore, the comparison focuses on parallel organization, state sharding, activation policy, and runtime scheduling, rather than low-level kernel differences.

We consider two model configurations: LLaMA-2-13B and Qwen2.5-32B, both with sequence length 2048, 256 MT-3000 compute clusters, global batch 4096, and 204.8M-token-equivalent training budget. Compared strategies include TP-heavy, ZeRO-3-heavy, Backward checkpointing (Ckpt), Full-save, Tuned PP/DP/ZeRO, and full RATrain. Tuned PP/DP/ZeRO allows searching P, D, Z, b, and A, but disables RATrain's layer-wise state pipeline, next-iteration update--prefetch scheduling, and FSR.

\begin{table*}[ht]
\centering
\scriptsize
\setlength{\tabcolsep}{2.5pt}
\renewcommand{\arraystretch}{0.86}
\caption{End-to-end comparison of GPU-style strategies on MT-3000.}
\label{tab:6p3_e2e}
\resizebox{\textwidth}{!}{%
\begin{tabular}{l l l c c c c}
\toprule
\textbf{Model} & \textbf{Method} & \textbf{Best Config} & \textbf{Peak Mem.(GB)} & \textbf{Step Time (s)} & \textbf{Tokens/s} & \textbf{Slowdown} \\
\midrule
LLaMA-2-13B & \textbf{RATrain} & \textbf{P=2,D=128,T=1,Z=2,b=1,A=32, FSR} & \textbf{15.84} & \textbf{688.09} & \textbf{12191.13} & \textbf{1.00$\times$} \\
LLaMA-2-13B & TP-heavy & P=2,D=64,T=2,Z=2,b=1,A=64, FSR & 16.51 & 826.53 & 10149.20 & 1.20$\times$ \\
LLaMA-2-13B & ZeRO-3-heavy & P=2,D=128,T=1,Z=3,b=1,A=32, FSR & 14.73 & 717.93 & 11684.48 & 1.04$\times$ \\
LLaMA-2-13B & Backward Ckpt & P=2,D=128,T=1,Z=2,b=1,A=32, Ckpt & 15.73 & 937.04 & 8952.21 & 1.36$\times$ \\
LLaMA-2-13B & Full-save & OOM & -- & -- & -- & -- \\
LLaMA-2-13B & Tuned PP/DP/ZeRO & P=2,D=128,T=1,Z=2,b=1,A=32, Ckpt & 15.73 & 945.84 & 8868.90 & 1.37$\times$ \\
\midrule
Qwen2.5-32B & \textbf{RATrain} & \textbf{P=8,D=32,T=1,Z=2,b=1,A=128, FSR} & \textbf{14.71} & \textbf{1592.51} & \textbf{5267.52} & \textbf{1.00$\times$} \\
Qwen2.5-32B & TP-heavy & P=8,D=16,T=2,Z=2,b=2,A=128, FSR & 19.45 & 1922.66 & 4363.01 & 1.21$\times$ \\
Qwen2.5-32B & ZeRO-3-heavy & P=8,D=32,T=1,Z=3,b=2,A=64, FSR & 16.54 & 1798.78 & 4663.50 & 1.13$\times$ \\
Qwen2.5-32B & Backward Ckpt & P=8,D=32,T=1,Z=2,b=1,A=128, Ckpt & 14.50 & 2162.54 & 3879.06 & 1.36$\times$ \\
Qwen2.5-32B & Full-save & OOM & -- & -- & -- & -- \\
Qwen2.5-32B & Tuned PP/DP/ZeRO & P=8,D=32,T=1,Z=2,b=1,A=128, Ckpt & 14.50 & 2167.81 & 3869.62 & 1.36$\times$ \\
\bottomrule
\end{tabular}}
\vspace{-0.5em}
\end{table*}

\begin{figure}[t]
\centering
\includegraphics[width=0.95\linewidth]{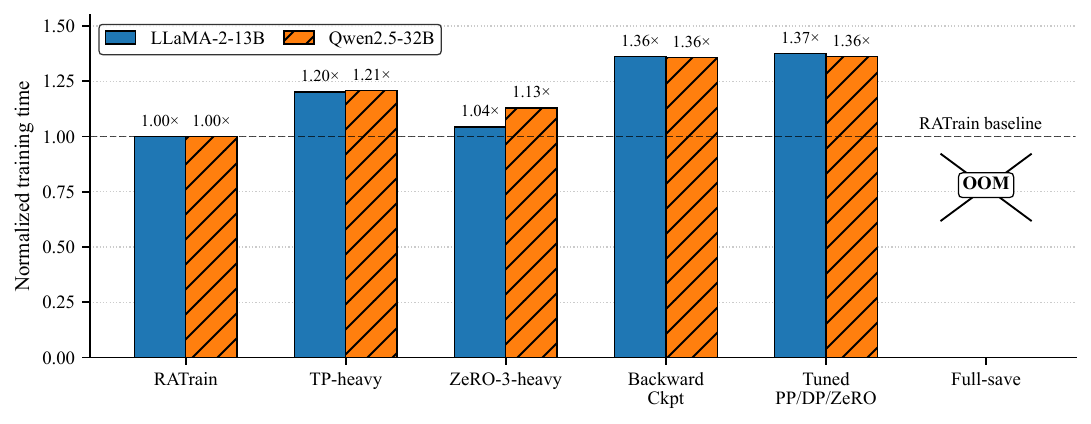}
\caption{Normalized step time for LLaMA-2-13B and Qwen2.5-32B, with RATrain as baseline. TP-heavy, ZeRO-3-heavy, Backward Ckpt, and Tuned PP/DP/ZeRO illustrate alternative GPU-style strategies. Full-save triggers OOM.}
\label{fig:6p3_normalized_training_time}
\end{figure}

Table~\ref{tab:6p3_e2e} shows that RATrain achieves the lowest step time on both LLaMA-2-13B and Qwen2.5-32B, reaching 12,191.13 and 5,267.52 tokens/s, respectively. RATrain chooses PP+DP+ ZeRO-2+FSR with T=1 in both models. Figure~\ref{fig:6p3_normalized_training_time} shows the RATrain-normalized step time, highlighting end-to-end efficiency gaps across GPU-style strategies.

Results indicate that, under MT-3000’s limited cross-cluster bandwidth, reducing intra-layer collectives is more critical than further splitting single-layer computation. TP-heavy is 1.20$\times$ and 1.21$\times$ slower than RATrain for the two models, because tensor parallelism reduces local compute but introduces intra-layer activation collectives and limits data-parallel degree. ZeRO-3-heavy is 1.04$\times$ and 1.13$\times$ slower, showing that aggressive state sharding adds extra parameter-view materialization and synchronization, while PP+ZeRO-2 already satisfies 20GB per-cluster memory. Full-save triggers OOM on both models, indicating full activation save is unsuitable for this memory budget.

Backward Ckpt uses the same PP/DP/ZeRO configuration as RATrain, but places activation recovery on the backward critical path, causing ~1.36$\times$ slowdown on both models. Tuned PP/DP/ZeRO demonstrates that tuning PP/DP/ZeRO alone cannot match RATrain: disabling layer-wise state pipeline, update--prefetch scheduling, and FSR yields step times close to Backward Ckpt. Overall, RATrain’s gains derive from combined parallel configuration and layer-level lifecycle scheduling, not a single mechanism; GPU-style empirical strategies cannot be directly applied to bandwidth-limited heterogeneous supercomputers.

\subsection{Resource-Constrained Training Capability}

This section evaluates RATrain's ability to support dense LLM training under the 20GB per-cluster memory constraint. This section examines whether the resource-aware planner can find the minimum executable resource configuration for each model. For each model, RATrain searches for the minimum number of clusters that satisfies the memory constraint, and then runs a 20.48M-token short validation under the selected configuration to verify resource feasibility.

Table~\ref{tab:min-feasible} reports the minimum feasible configurations at sequence length 2048. RATrain supports dense LLM training from LLaMA-2-7B to LLaMA-2-70B under the 20GB per-cluster memory constraint. As model size increases, the planner gradually increases the pipeline degree to reduce per-stage residency of parameters, activations, and optimizer states: LLaMA-2-7B uses $P=2$, Baichuan2-13B uses $P=8$, Qwen2.5-32B uses $P=16$, and LLaMA-2-70B uses $P=48$. This trend indicates that pipeline parallelism is the primary mechanism for scaling model size under strict per-cluster memory constraints.
\begin{table}[t]
\centering
\scriptsize
\setlength{\tabcolsep}{2.6pt}
\renewcommand{\arraystretch}{0.90}
\caption{Minimum feasible training configurations under the 20GB per-cluster memory constraint.}
\label{tab:min-feasible}
\resizebox{\linewidth}{!}{%
\begin{tabular}{lclccc}
\toprule
\textbf{Model} & \textbf{Min.} & \textbf{Config} & \textbf{Peak Mem.} & \textbf{Step Time} & \textbf{Tokens/s} \\
 & \textbf{Clusters} & & \textbf{(GB)} & \textbf{(s)} & \\
\midrule
LLaMA-2-7B    & 8  & $P=2,D=4,A=128$   & 19.57 & 1304.13 & 804.04 \\
Baichuan2-13B & 16 & $P=8,D=2,A=128$   & 19.06 & 743.15  & 705.50 \\
Qwen2.5-32B   & 64 & $P=16,D=4,A=128$  & 18.14 & 873.85  & 1199.96 \\
LLaMA-2-70B   & 96 & $P=48,D=2,A=16$   & 19.46 & 281.32  & 232.96 \\
\bottomrule
\end{tabular}%
}
\end{table}
All configurations in Table~\ref{tab:min-feasible} keep $T=1$ and use PP+DP+ ZeRO-2+FSR. This is consistent with the observation in Section~\ref{sec:eval-gpustyle}: on a platform with limited inter-cluster communication bandwidth, avoiding intra-layer tensor-parallel collectives is often more effective than further splitting single-layer computation. Meanwhile, ZeRO-2 already provides sufficient state partitioning for these configurations, avoiding the additional parameter-view materialization and synchronization overheads introduced by ZeRO-3.

Overall, RATrain's planner adapts pipeline degree and data-parallel degree according to model size and memory constraints, while using FSR to control activation residency. Even for LLaMA-2-70B, RATrain completes short validation on 96 MT-3000 compute clusters, demonstrating that it can provide a practically executable training configuration for 70B-class dense LLMs under the 20GB per-cluster memory constraint.

\subsection{Planner Accuracy}

RATrain's resource-aware planner uses model, platform, and execution profiles to select a training configuration that satisfies the memory constraint and minimizes the predicted step time. This section evaluates the planner's prediction accuracy by comparing its predicted step time with the measured execution time on MT-3000.

\begin{table}[t]
\centering
\scriptsize
\setlength{\tabcolsep}{3.0pt}
\renewcommand{\arraystretch}{0.90}
\caption{Planner prediction accuracy on representative configurations.}
\label{tab:planner-accuracy}
\resizebox{\linewidth}{!}{%
\begin{tabular}{lcccc}
\toprule
\textbf{Model} & \textbf{Clusters} & \textbf{Pred. Step (s)} & \textbf{Meas. Step (s)} & \textbf{Error} \\
\midrule
LLaMA-2-7B    & 256 & 140.92 & 144.28 & 2.33\% \\
Baichuan2-13B & 256 & 268.74 & 276.61 & 2.85\% \\
Qwen2.5-32B   & 256 & 441.83 & 455.21 & 2.94\% \\
Qwen2.5-32B   & 512 & 225.47 & 231.36 & 2.55\% \\
\bottomrule
\end{tabular}%
}
\end{table}

Table~\ref{tab:planner-accuracy} reports the prediction results under representative model sizes and resource budgets. After the planner selects a configuration according to its cost model, we run the selected configuration on the real MT-3000 backend and measure its actual step time. The prediction error ranges from 2.33\% to 2.94\%, with an average error of 2.67\%. These results indicate that RATrain's execution profiles capture the dominant execution costs across different model sizes and resource scales, thereby reducing the configuration search overhead. All subsequent end-to-end performance, ablation, and scalability results are based on measurements on the MT-3000 platform.

\subsection{Sequence-Length Sensitivity and Compute Utilization}

This section evaluates the impact of sequence length on RATrain's training time and compute utilization. Sequence length changes the attention computation, activation residency, checkpoint/recovery cost, and stage-local memory pressure, and therefore serves as an important dimension for testing whether RATrain is tuned only for a fixed input length. The experiments are conducted on 256 MT-3000 compute clusters with a global batch size of 4096, covering sequence lengths of 512, 1024, 2048, 3072, and 4096. The evaluated models include LLaMA-2-7B, Baichuan2-13B, and Qwen2.5-32B. Training time is normalized to the time required to process 204.8M tokens, and compute utilization is reported using a MAC-only metric.
\begin{figure}[t]
  \centering
  \includegraphics[width=\linewidth]{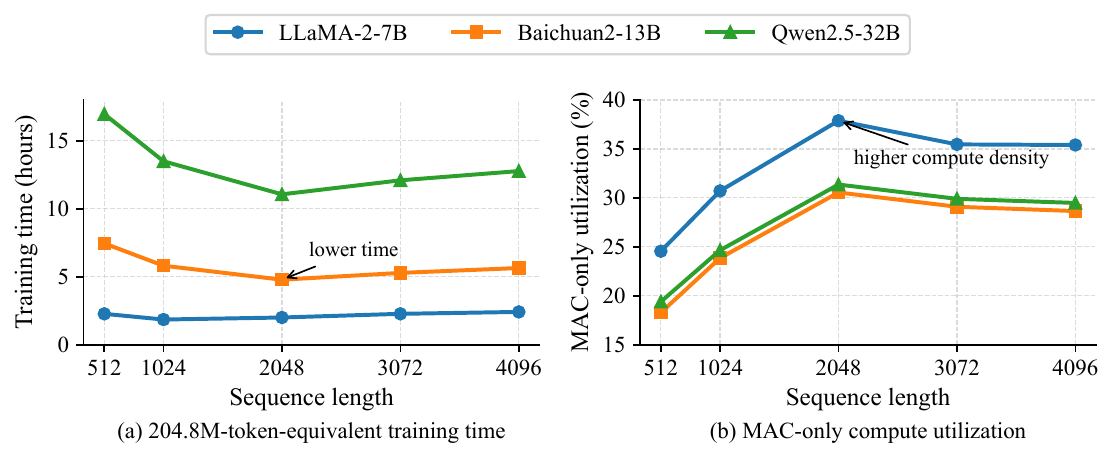}
  \caption{
  Sequence-length sensitivity of RATrain.
  (a) 204.8M-token-equivalent training time under different sequence lengths.
  (b) MAC-only compute utilization under different sequence lengths.
  Experiments use 256 MT-3000 compute clusters and global batch size 4096.
  }
  \label{fig:seq-sensitivity}
\end{figure}

Figure~\ref{fig:seq-sensitivity} shows the training time and MAC-only utilization under different sequence lengths. Overall, the training time of Baichuan2-13B and Qwen2.5-32B decreases from 512 to 2048 and increases again at longer sequences; LLaMA-2-7B performs well around 1024 and 2048. The MAC-only utilization of all three models increases from 512 to 2048 and slightly decreases at 3072 and 4096. This indicates that medium sequence lengths better amortize fixed scheduling, communication, and state-task overheads, while overly long sequences increase attention, activation, and recovery pressure.

\begin{table}[t]
\centering
\scriptsize
\setlength{\tabcolsep}{3.0pt}
\renewcommand{\arraystretch}{0.90}
\caption{Representative FP16 GEMM backend profile at sequence length 2048.}
\label{tab:gemm-profile}
\resizebox{\linewidth}{!}{%
\begin{tabular}{lccc}
\toprule
\textbf{GEMM Shape} & \textbf{MAC Util. (\%)} & \textbf{Throughput (T MAC/s)} & \textbf{Latency (ms)} \\
\midrule
4096$\times$4096   & 64.96 & 5.26 & 6.53 \\
4096$\times$11008  & 66.16 & 5.36 & 17.23 \\
11008$\times$4096  & 65.13 & 5.28 & 17.50 \\
6656$\times$6656   & 67.35 & 5.46 & 16.63 \\
8192$\times$8192   & 68.13 & 5.52 & 24.90 \\
\bottomrule
\end{tabular}%
}
\end{table}

To explain the backend basis of utilization, Table~\ref{tab:gemm-profile} reports representative FP16 GEMM profiles at sequence length 2048. RATrain's GEMM backend maintains 64.96\%--68.13\% MAC utilization on projection, FFN, and larger hidden-size square GEMMs, corresponding to 5.26--5.52 T MAC/s effective throughput. The end-to-end MAC-only utilization is lower than the per-GEMM utilization mainly because a full training step also includes non-GEMM overheads, such as attention, activation recovery, state synchronization, parameter update, and data movement.

\begin{figure}[t]
\centering
\includegraphics[width=\linewidth]{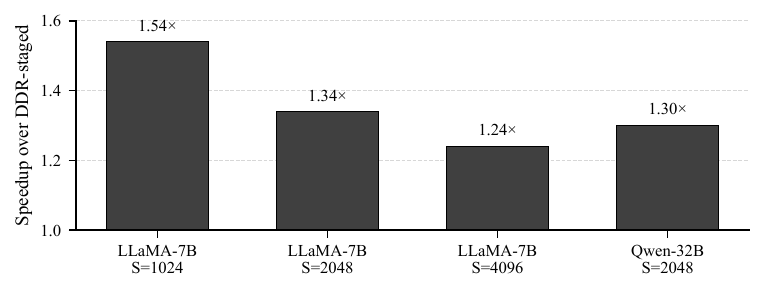}
\caption{Memory-resident Attention BP speedup over the DDR-staged baseline.}
\label{fig:attn-bp-speedup}
\end{figure}

Beyond GEMM, Attention BP is the part of backward computation most sensitive to memory access and intermediate-state movement. Figure~\ref{fig:attn-bp-speedup} shows that memory-resident Attention BP consistently outperforms the DDR-staged baseline: it achieves 1.54$\times$, 1.34$\times$, and 1.24$\times$ speedups on the LLaMA-2-7B layer at sequence lengths 1024, 2048, and 4096, respectively, and 1.30$\times$ on the Qwen2.5-32B layer at sequence length 2048, with an average speedup of 1.36$\times$. This shows that the memory-resident tile schedule reduces DDR traffic for intermediate states in backward computation.

Overall, RATrain maintains executability and stable efficiency across different input lengths and model scales. Sequence length 2048 generally provides a favorable balance between compute density and memory pressure, but RATrain does not rely on a fixed sequence length. Instead, it uses resource-aware planning to balance compute density, activation residency, and recovery overhead.

\subsection{Ablation Study}

This section analyzes the contribution of RATrain's core mechanisms to training performance. The experiment uses Qwen2.5-32B with sequence length 2048, 256 MT-3000 compute clusters, and global batch size 4096. All variants use the same configuration, $P=8,D=32,T=1,Z=2,b=1,A=128$, and only disable the corresponding mechanism under study. Full RATrain achieves a step time of 1790.13 s with an exposed tail of 14.69 s.
\begin{figure}[t]
  \centering
  \includegraphics[width=\linewidth]{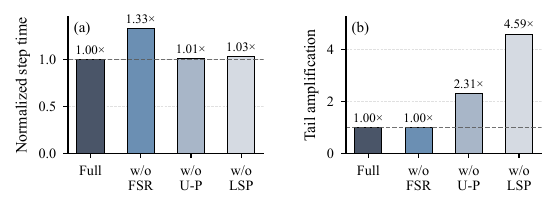}
  \caption{
  Ablation study on Qwen2.5-32B.
  Step time and exposed tail are normalized to Full RATrain.
  U-P denotes update--prefetch scheduling, and LSP denotes layer-wise state pipeline.
  }
  \label{fig:ablation}
\end{figure}
Figure~\ref{fig:ablation} shows the normalized step time and exposed-tail amplification. FSR has the largest impact on end-to-end step time: removing FSR increases the step time to 1.33$\times$. This indicates that backward-time recomputation exposes activation recovery on the backward critical path, whereas RATrain's FSR moves part of the recovery cost into pipeline-side windows.

The two state-scheduling mechanisms mainly reduce the exposed tail. Disabling update--prefetch scheduling increases tail amplification to 2.31$\times$, indicating that next-iteration parameter-view preparation can cause readiness stalls before forward execution. Disabling the layer-wise state pipeline further increases tail amplification to 4.59$\times$, showing that gradient synchronization and the subsequent state update and parameter preparation tasks can form a significant state-processing tail if they are concentrated near the step boundary.

Overall, the ablation results show that RATrain's gains do not come from a single optimization, but from the coordinated scheduling of activation recovery, state lifecycle management, and next-iteration parameter readiness.

\subsection{Resource Scalability}

This section evaluates RATrain's throughput scalability as the number of MT-3000 compute clusters increases. The experiment uses LLaMA-2-7B with sequence length 2048, and scales the number of compute clusters from 256 to 512, 768, and 1024. This section adopts a throughput-oriented scale-out setting: the local training configuration of each data-parallel replica is kept unchanged, while the data-parallel degree and global batch size are increased linearly with the number of clusters. Therefore, this experiment focuses on whether RATrain can translate additional resources into higher training throughput, rather than fixed-global-batch strong scaling.

\begin{table}[t]
\centering
\scriptsize
\setlength{\tabcolsep}{3.0pt}
\renewcommand{\arraystretch}{0.90}
\caption{Throughput-oriented scale-out results on LLaMA-2-7B.}
\label{tab:resource-scalability}
\resizebox{\linewidth}{!}{%
\begin{tabular}{cccccc}
\toprule
\textbf{Clusters} & \textbf{Global Batch} & \textbf{Step Time} & \textbf{Tokens/s} & \textbf{Speedup} & \textbf{Efficiency} \\
 & & \textbf{(s)} & & & \\
\midrule
256  & 2048 & 144.28 & 29,069.73  & 1.00$\times$ & 100.0\% \\
512  & 4096 & 145.75 & 57,558.07  & 1.98$\times$ & 99.0\% \\
768  & 6144 & 147.23 & 85,465.01  & 2.94$\times$ & 98.0\% \\
1024 & 8192 & 148.75 & 112,790.55 & 3.88$\times$ & 97.0\% \\
\bottomrule
\end{tabular}%
}
\end{table}

Table~\ref{tab:resource-scalability} reports the scale-out results. As the number of compute clusters increases from 256 to 1024, the step time only increases from 144.28 s to 148.75 s, while throughput improves from 29,069.73 tokens/s to 112,790.55 tokens/s, corresponding to a 3.88$\times$ speedup. Scaling efficiency decreases from 100.0\% to 97.0\%, mainly due to gradient synchronization, runtime scheduling, and system variability at larger data-parallel group sizes. Overall, these results show that RATrain can effectively convert additional MT-3000 compute clusters into higher training throughput while keeping the local training path stable.
\section{Related Work}

\textbf{Distributed LLM training systems.}
Existing large-model training systems are mainly designed for GPU clusters, and scale Transformer training through tensor parallelism, pipeline parallelism, data parallelism, and state partitioning~\cite{shoeybi2019megatron,narayanan2021megatron,rajbhandari2020zero,rasley2020deepspeed}. Megatron-LM combines tensor and pipeline parallelism to train multi-billion-parameter models~\cite{shoeybi2019megatron,narayanan2021megatron}; GPipe, PipeDream, and PipeDream-2BW study pipeline scheduling, bubbles, activation storage, and weight versioning~\cite{huang2019gpipe,narayanan2019pipedream,narayanan2021pipedream2bw}; DeepSpeed and ZeRO reduce data-parallel state redundancy by partitioning optimizer states, gradients, and parameters~\cite{rajbhandari2020zero,rasley2020deepspeed}. These systems typically assume high-bandwidth device memory, high-speed interconnects, and mature collective libraries, whereas RATrain targets the MT-3000 heterogeneous supercomputing platform with explicit memory hierarchy, limited per-cluster memory, and bandwidth-constrained inter-cluster communication~\cite{lu2022mt}. RATrain therefore does not treat tensor parallelism or ZeRO-3 as the default scaling path, but schedules training-state lifecycles around PP/DP/lightweight-ZeRO and standard 1F1B execution.

\textbf{Automatic parallelism and configuration search.}
Automatic parallelism systems generate distributed execution plans through search or compilation-based optimization~\cite{flexflow2019,alpa2022,gspmd2021,whale2022}. FlexFlow searches parallelization strategies across operator, sample, attribute, and parameter dimensions~\cite{flexflow2019}; Alpa combines inter-operator and intra-operator parallelism for large-scale Transformer models~\cite{alpa2022}; GSPMD provides a general SPMD partitioning abstraction for tensor sharding and device mapping~\cite{gspmd2021}; Whale introduces a hardware-aware parallel strategy for heterogeneous GPU clusters~\cite{whale2022}. Unlike these systems, which mainly focus on computation graph partitioning, tensor sharding, and device placement, RATrain's planner explicitly models peak memory, exposed communication, activation recovery, parameter prefetching, and step-end state-processing tail, and directly materializes the selected plan as stage-local runtime tasks.

\textbf{Activation memory optimization and rematerialization.}
Activation checkpointing and tensor rematerialization reduce training memory by trading additional computation for lower activation residency~\cite{chen2016checkpoint,checkmate2020,capuchin2020}. Classical checkpointing recomputes missing activations during backward propagation~\cite{chen2016checkpoint}; Checkmate formulates rematerialization as an optimization problem between memory and recomputation cost~\cite{checkmate2020}; Capuchin combines tensor eviction, prefetching, and recomputation for GPU memory management~\cite{capuchin2020}; recent work further studies efficient activation rematerialization and bubble-filling transformations to reduce recomputation overhead in LLM training~\cite{yuan2024activation,huang2025obscura}. RATrain is complementary to these techniques: instead of only choosing checkpoint placement, it keeps the standard 1F1B order unchanged while moving recoverable activation reconstruction into stage-local windows before backward arrival, thereby reducing backward-path exposure.

\textbf{Heterogeneous training and offloading systems.}
Prior work also studies large-model training with heterogeneous memory or devices. ZeRO-Offload moves optimizer states and part of the computation to CPU to reduce GPU memory pressure~\cite{zerooffload2021}; Whale studies automatic parallelism and load balancing for heterogeneous GPUs~\cite{whale2022}; other systems extend trainable model scale through offloading, eviction, prefetching, or recomputation~\cite{zerooffload2021,capuchin2020,yuan2024activation}. RATrain targets a different class of platforms: heterogeneous supercomputers organized around autonomous compute clusters, software-managed memory hierarchy, and bandwidth-constrained inter-cluster communication~\cite{lu2022mt}. It does not use a slower memory tier merely as a GPU-memory extension; instead, it treats parameters, gradients, optimizer states, activations, and communication buffers as runtime objects with explicit lifecycles, and schedules them at stage-local and layer-level granularity.

\section{Conclusion}

This paper presents RATrain, a resource-aware training runtime for dense LLMs on bandwidth-constrained heterogeneous supercomputing platforms. RATrain models standard 1F1B training as a training-state lifecycle scheduling problem, and reduces exposed overhead through layer-level, stage-local scheduling of state synchronization, parameter preparation, and activation recovery, without changing training semantics. Experiments on a real MT-3000 platform show that RATrain can support 7B--72B dense LLM training configurations and achieves stable behavior in correctness, performance, mechanism effectiveness, and resource scalability.

\bibliographystyle{ACM-Reference-Format}
\bibliography{ratrain}

\end{document}